\documentstyle[12pt]{article}

\def\be{\begin{equation}}
\def\ee{\end{equation}}
\def\bea{\begin{eqnarray}}
\def\eea{\end{eqnarray}}
\def\vk{{\bf k}}
\def\vko{{\bf k}_{1}}
\def\vkt{{\bf k}_{2}}

\begin{document}

\begin{titlepage}

\begin{centering}
\vfill

{\bf  ESTIMATION OF THE PARTICLE-ANTIPARTICLE CORRELATION EFFECT \\
FOR PION PRODUCTION IN HEAVY ION COLLISIONS}
 
\vspace{2cm}

I.V.ANDREEV
\vspace{0.3cm}
{\em P.N.Lebedev Physical Institute, Moscow 117924, Russia}

\vspace{3cm}

{\bf Abstract}

\end{centering}

\vspace{0.3cm}\noindent
Estimation of the back-to-back $\pi-\pi$ correlations arising due to evolution
of the pionic field in the course of pion production process is given for
central heavy nucleus collisions at moderate energies.

\vfill \vfill

\end{titlepage}

\section{Introduction}

It is usually suggested that in high-energy heavy nucleus collisions
an excited volume is formed which undergoes evolution and subsequent decay
into free final particles. Particles existing in the excited volume represent
a part of the medium being rather quasiparticles than free particles. 
So the form of their energy spectrum $E_{\vk}$ may differ essentially
from that of free particles $E^{0}_{\vk}=({\vk}^{2}+m^{2})^{1/2}$.
It was noted (see~\cite{{AW},{AC},{HM},{A}}) that this feature leads to some
modification of the well known identical particle correlations (HBT effect)
and also to appearance of specific back-to-back particle-antiparticle
correlations (PAC effect). No practical estimations of the pionic PAC effect
is known to us.

Below we consider central heavy nocleus collisions at moderate (up to AGS)
energies. In this case the excited volume consists mainly from nucleons
and pions (at least at the late stage). We consider final state pion
correlations. The PAC effect is determined through the evolution parameter
$r({\vk})$ and depends on two factors: first, to what extent the pionic
energy $E_{\vk}$ is modified and second, what is the characteristic time
$t_{0}$ of the energy spectrum evolution. Our numerical estimations showed
that the pion modification in the course of the hadronic matter evolution
(say expansion, cooling) is too slow to give sizable PAC effect. So we
consider only the fast breakup of the hadronic matter into free pions
(freezeout) as a source of PAC. Usually the breakup is considered as an
instantaneous process (neglecting its time duration $t_0$) thus ensuring
maximal PAC effect. However the PAC effect under consideration is sensitive
to rather small time intervals of the order of $1  fm$. So below we estimate
PAC for finite $t_0$.

\section{HBT and PAC effects}

In this section we describe in parallel the main features of PAC and HBT
correlations taking into account pionic energy modification. To present the
results in a simple form we use here a simplified description of the excited
hadronic volume (particle source). So the following expressions represent
a limiting case of those in Refs.~\cite{AW,A}
 \footnote{
 The sign of one of two
momenta ${\bf p}_{1},{\bf p}_{2}$ in the right hand side of the second of
Eqs.14 in the Ref.~\cite{AW} must be changed. Then Eqs.13-14 of that paper
will be applicable for neutral pions. This erroneous sign appeared because
of neglect of the difference between relativistic quantum field and a simple
set of quantum oscillators in Ref.~\cite{AW}.
}.
The volume is suggested to be homogenious, motionless (neglect of the flow),
isotopically symmetric
and large enough (heavy nuclei). Under these conditions the sigle-particle
inclusive cross-section can be written in a simple form:
\bea
N(\vk)=\frac{1}{\sigma}\frac{d\sigma}{d^{3}k}=<a^{\dag}(\vk)a(\vk)>
=\frac{V}{(2\pi)^3}\left[ n(\vk)+(2n(\vk)+1)\sinh^{2}r(\vk)\right]
\label{eq:1}
\eea
where $a^{\dag},a$ are creation and annihilation operators of the final state
pions, $V$ is the volume of the source, $n(\vk)$ is the level occupation number
(for example, Bose distribution) and $r(\vk)$ is the evolution parameter.

Two-particle inclusive cross-sections are given by
\bea
\frac{1}{\sigma}\frac{d^{2}\sigma^{++}}{d^{3}k_{1}d^{3}k_{2}}
=\langle a^{\dag}_{1}a^{\dag}_{2}a_{1}a_{2}\rangle
=\langle a^{\dag}_{1}a_{1}\rangle\langle a^{\dag}_{2}a_{2}\rangle
+\langle a^{\dag}_{1}a_{2}\rangle\langle a^{\dag}_{2}a_{1}\rangle
\label{eq:2}
\eea
for like sign charged (identical) pions (HBT effect),
\bea
\frac{1}{\sigma}\frac{d^{2}\sigma^{+-}}{d^{3}k_{1}d^{3}k_{2}}
=\langle a^{\dag}_{1}b^{\dag}_{2}a_{1}b_{2}\rangle
=\langle a^{\dag}_{1}a_{1}\rangle\langle b^{\dag}_{2}b_{2}\rangle 
+\langle a^{\dag}_{1}b^{\dag}_{2}\rangle\langle a_{1}b_{2}\rangle
\label{eq:3}
\eea
for charged particle-antiparticle $(\pi^{+}\pi^{-})$ pairs (PAC) and
\bea
\frac{1}{\sigma}\frac{d^{2}\sigma^{00}}{d^{3}k_{1}d^{3}k_{2}}
=\langle a^{\dag}_{1}a^{\dag}_{2}a_{1}a_{2}\rangle
=\langle a^{\dag}_{1}a_{1}\rangle\langle a^{\dag}_{2}a_{2}\rangle
+\langle a^{\dag}_{1}a_{2}\rangle\langle a^{\dag}_{2}a_{1}\rangle
+\langle a^{\dag}_{1}a^{\dag}_{2}\rangle\langle a_{1}a_{2}\rangle
\label{eq:4}
\eea
for neutral pion pairs (both HBT and PAC) with
\be
\langle a^{\dag}(\vko)a(\vkt)\rangle
=\left[ n(\vk)+(2n(\vk)+1)\sinh^{2}r(\vk)\right]F(\vko-\vkt)
\label{eq:5}
\ee
\be
\langle a(\vko)b(\vkt)\rangle
=\sinh2r(\vk)\left[n(\vk)+\frac12\right] F(\vko+\vkt)
\label{eq:6}
\ee
(the same for $<a(\vko)a(\vkt)>$ in the case of neutral pions).
In Eqs.5-6 $r(\vk)$ is the evolution parameter and the function $F(\vko\pm\vkt)$
represents the Fourier transform of the source volume at breakup stage.
This is sharply peaked function of $\vko\pm\vkt$ (at zero momentum) having
characteristic scale of the order of inverse size of the source, this scale
being much less than characteristic scales of pion momentum distribution
$n(\vk)$ and evolution parameter $r(\vk)$. So the last two functions may be
evaluated at any of momenta $\vko,\vkt\approx\pm\vk$ (we suggest that
the process is $\vk\to -\vk$ symmetric, for example collision of identical
nuclei at CMS). Evidently pion-pion interaction effects are not present in
Eqs.2-4; it is supposed (as usual) that they can be separated from exposed
HBT-type correlations.

Relative two-particle correlation functions which are measured in experiment
are given by
\be
C^{++}_{2}(\vko,\vkt)=1+\frac{|<a^{\dag}(\vko)a(\vkt)>|^{2}}{N(\vko)N(\vkt)}
\label{eq:7}
\ee
\be
C^{+-}_{2}(\vko,\vkt)=1+\frac{|<a(\vko)b(\vkt)>|^{2}}{N(\vko)N(\vkt)}
\label{eq:8}
\ee
\be
C^{00}_{2}(\vko,\vkt)=1+\frac{|<a^{\dag}(\vko)a(\vkt)>|^{2}}{N(\vko)N(\vkt)}
+\frac{|<a(\vko)a(\vkt)>|^{2}}{N(\vko)N(\vkt)}
\label{eq:9}
\ee
Introducing normalized form-factor of the pre-breakup volume,
\be
F(\vko\pm\vkt)=\frac{V}{(2\pi)^{3}}G(\vko\pm\vkt), \qquad G(0)=1
\label{eq:10}
\ee
we therefore get
\be
C^{++}(\vko,\vkt)=1+G^{2}(\vko-\vkt)
\label{eq:11}
\ee
\be
C^{+-}(\vko,\vkt)=1+c(\vk)G^{2}(\vko+\vkt)
\label{eq:12}
\ee
\be
C^{00}(\vko,\vkt)=1+G^{2}(\vko-\vkt)+c(\vk)G^{2}(\vko+\vkt)
\label{eq:13}
\ee
with
\be
c(\vk)=\left(\frac{\sinh r(\vk)\cosh r(\vk)(2n(\vk)+1)}
                  {\sinh^{2}r(\vk)(2n(\vk)+1)+n(\vk)}\right)^{2}
\label{eq:14}
\ee
where Eq.11 gives the usual HBT effect and Eq.12 describes the PAC effect.
Both of them contain the same form-factor $G(\vk)$ (in our approximation)
ensuring the same direction $\pi^{+}\pi^{+}$ correlations and back-to-back
$\pi^{+}\pi^{-}$ correlations. Neutral pions show both kinds of the
correlations being identical particles and simultaneously antiparticles
to themselves.

As it can be seen from Eq.14 the PAC effect is determined through evolution
parameter $r(\vk)$. In turn the evolution parameter depends on time duration
$t_0$ of the pion energy evolution. For very small characteristic times $t_0$
the expression for $r(\vk)$ is universal~\cite{AW},
\be
r_{m}=\frac12\ln\left( \frac{E^{0}_{\vk}}{E_{\vk}}\right),\qquad t_{0}=0
\label{eq:15}
\ee
where $E_{\vk}$ is the pion energy at pre-breakup moment and $E^{0}_{\vk}$
is free pion energy. For larger $t_0$ the evolution parameter lessens and
depends on unknown details of the breakup process. For estimation of $r(\vk)$
we shall use the model expression of Ref.~\cite{A}:
\be
r(\vk)=\frac12\ln\left(\frac{\tanh(\pi E^{0}_{\vk}t_{0}/2)}
                         {\tanh(\pi E_{\vk}t_{0}/2)}\right)
\label{eq:16}
\ee
Below we estimate the evolution parameter $r(\vk)$ and the factor $c(\vk)$
in Eq.14 which determines the strength of the PAC effect.

\section{Estimation of PAC in finite nucleon density matter}
To evaluate PAC one has to know the pion energy spectrum $E_{\vk}$ in finite
nucleon density matter. The simplest way to find the energy spectrum is
the use of the notion of the pseudopotential~\cite{GW}. It is determined
as an effective potential in which the pions propagate and it is given
by the forward scattering amplitude $f(\vk)$ of the pions on the particles
of the medium,
\be
U(\vk)=-4\pi\rho<f(\vk)>
\label{eq:17}
\ee
where $\rho$ is the density of the medium particles and the amplitude $f(\vk)$
is averaged over states of the medium particles.

In finite nucleon density matter $f(\vk)$ is mainly $\pi N$ scattering
amplitude. The main contribution to the amplitude is given here by P-wave
scattering dominated by delta resonance. Corresponding momentum dependent
effective potential for isotopically symmetric (number of protons is equal
(close) to number of neutrons) matter may be taken in the form
(see also Refs.~\cite{M,OTW}):
\be
U(\vk)=-\frac89 f_{\Delta}^{2}\frac{M_{\Delta}E_{\Delta}}{Mm^{2}}
\frac{\vk^{2}}{E_{\Delta}^{2}-\vk^{2}-m^{2}}
\label{eq:18}
\ee
with
\be
f_{\Delta}^{2}/4\pi=0.37
\label{eq:19}
\ee
\be
E_{\Delta}=(M_{\Delta}^{2}-M^{2}-iM_{\Delta}\Gamma_{\Delta})\left/2M\right.
=(2.4-0.5i)m
\label{eq:20}
\ee
where $f_{\Delta}$ is empiric $\pi N\Delta$ coupling constant, $m$ is pion
mass, $M$ is nucleon mass, $M_{\Delta}$ and $\Gamma_{\Delta}$ are mass and
width of the delta resonance. Eqs.18-20 represent the sum of the direct
and exchange $\pi N$ scattering diagrams with delta resonance in the
intermediate state where we neglected nucleon velocities and omitted
terms containing inverse nucleon mass (first order $M^{-1}$-terms give
only a few per cent correction to $E_{\Delta}$). The pion energy in the
nucleon matter is now given by the equation
\be
E^{2}_{\vk}=m^{2}+\vk^{2}+U(\vk)
\label{eq:21}
\ee
It is shown at Fig.1 for nucleon density $\rho$ equal to one half of the
nuclear matter density (energies and momenta are taken in pion mass units).


Let us note that Eq.21 has the form of the pionic dispersion equation with
substitution of the effective potential $U(\vk)$ for pionic polarization
operator $\Pi(\vk,E_{\vk})$ which depends both on momentum $\vk$ and energy
$E_{\vk}$. The polarization operator $\Pi(\vk,E_{\vk})$ in the same
approximation is given by the Eq.18 with substitution of the energy squared
$E_{\vk}^2$ for $\vk^{2}+m^2$ in the denominator of the Eq.18 and the pion
spectrum (excitations having pion quantum numbers) is then given by
selfconsistent solution of the resulting dispersion equation. At first sight
the resulting pionic energy spectrum~\cite{M} differs essentially from that
of Eq.21, containing at least two branches shown at Fig.1 by dashed lines
(original pion and delta-hole mixed states tending to be intersecting ones
in the limit of zero coupling constant $f_{\Delta}$). However, considering 
effects of the pion energy evolution one must use two pieces of these two
branches which correspond to true pion and we return essentially to 
single branch given by Eq.21, see Fig.1. The vicinity of the would-be
intersection point (the resonant point $\vk^{2}_{res}+m^{2}=Re E^{2}_{\Delta},
\quad k_{res}=2.1m$, where these two descriptions still differ, does not contribute
in any case (here $r(k_{res})=0$ and the imaginary part of the pion energy
is maximal and large). All that justifies the use of the pseudopotential 
$U(\vk)$ for calculation of the evolution parameter $r(\vk)$ (introduction
of the polarization operator $\Pi(\vk,E_{\vk})$ would require a reformulation
of the scheme of calculation of the evolution effects).

To evaluate the evolution effects it is necessary to specify the level
population $n(\vk)$, the nucleon density $\rho$ at breakup and the time
duration $t_0$ of the breakup stage of the process. The level population
may be approximated by Bose distribution with empiric temperature which
we take to be equal to $120 MeV$, 
\be
n(\vk)=\left( \exp(E_{\vk}/T)-1\right)^{-1}, \qquad T=120 MeV
\label{eq:22}
\ee
It seems reasonable to take the nucleon density $\rho$ to be slightly less
than nuclear matter density $\rho_{n}$. So below we present estimations 
for two values of $\rho$ which we consider as limiting ones, $\rho=0.5\rho_{n}$
and $\rho=\rho_{n}$. The time duration $t_{0}$ is left as a free parameter.    

One more problem is the permissible range of the the pion momenta $\vk$ where
the potential $U(\vk)$ given by Eq.18 (that is corresponding scattering
amplitude) may be used. Evidently this is a low-energy potential applicable
at the most at $k\le (3-4)m$. Furthermore the imaginary part of the
potential must not exceed the difference between quasipion energy and free
pion energy. This leaves us small momentum region $k\le1.5m$ where the
calculation of the PAC effect seems to be reliable. It must be also noted that
just above the delta resonant energy ($k_{res}=2.1m$) there is another source
of back-to-back pairs --  $\rho$ meson decay ($k=(2.5\pm 0.5)m$ for
free $\rho$ mesons). So the PAC effect under consideration is an additional
possible source of the correlated $\pi\pi$ pairs active at lower energies.

Calculation of the evolution parameter $r(\vk)$ according to Eqs.15-16
together with Eqs.19-21 shows that in the case under consideration it is
rather small being zero at $k=0$ and at $k=k_{res}=2.1m$ and reaching
the maximal values:\\
 $r_{max}(\vk)=0.16$ at $k=1.6m$ for $\rho=0.50\rho_{n}$, $t_{0}=0$;\\
 $r_{max}(\vk)=0.22$ at $k=1.5m$ for $\rho=0.75\rho_{n}$, $t_{0}=0$;\\
 $r_{max}(\vk)=0.30$ at $k=1.3m$ for $\rho=1.00\rho_{n}$, $t_{0}=0$.\\
The function $r(\vk)$ decreases fastly when characteristic time $t_{0}$
increases and vanishes at $t_{0}\approx 2 fm$.

The enhancement of single-particle inclusive cross-sections arising due to
evolution effect is shown at Fig.2a,b where the distribution $N(\vk)$, given by
Eq.1, over the unenhanced ($r=0$) value $N_{0}(\vk)$ is depicted for different
values of the characteristic time $t_{0}$ for two nucleon densities
$\rho=0.5\rho_{n}$ and $\rho=\rho_{n}$ (pion momenta at Figs.2,3 are taken 
in pion mass units).


Corresponding results for the factor $c(\vk)$ in Eqs.12-14, which gives the
strength of the PAC effect, are shown at Fig.3a,b.


As can be seen from Fig.3 the PAC effect can be essential if the characteristic
breakup time $t_{0}$ is small enough ($t_{0}<1 fm$). The presence or absence
of PAC can serve as a measure of the time duration $t_{0}$ about which we
have no other information. Contrary to single-particle enhancement, which can
have various origin, the PAC effect (if any) can be unambiguously identified
through measurement of the excess of say zero rapidity (in CMS) small
momentum correlated particle-antiparticle pairs.

\section{Conclusions}
Estimation of the pionic PAC effect in heavy nucleus collisions shows that
it can serve as a substantial source of back-to-back $\pi^{+}\pi^{-}$
and $\pi^{0}\pi^{0}$ pairs (ensuring also an enhancement of single-particle
pion distributions) if breakup (freezout) time is small enough.

\section*{Figure captions}
Fig.1 Quasipion energy in nucleon medium (with $\rho=0.5\rho_{n}$) calculated
      through pseudopotential (solid line) and polarization operator (dashed
      lines) together with free pion energy (dotted line).\\
Fig.2 Enhancement of single-particle inclusive cross-section due to evolution
      effect for $t_{0}=0,\;0.5,\;1.0,\;1.5fm$ (from top to bottom).
      a)$\rho=0.5\rho_{n}$, b)$\rho=\rho_{n}$.\\
Fig.3 Relative strength of PAC (Eqs.12-14) for $t_{0}=0,\;0.5,\;1.0,\;1.5 fm$
      (from top to bottom). a)$\rho=0.5\rho_{n}$, b)$\rho=\rho_{n}$.   

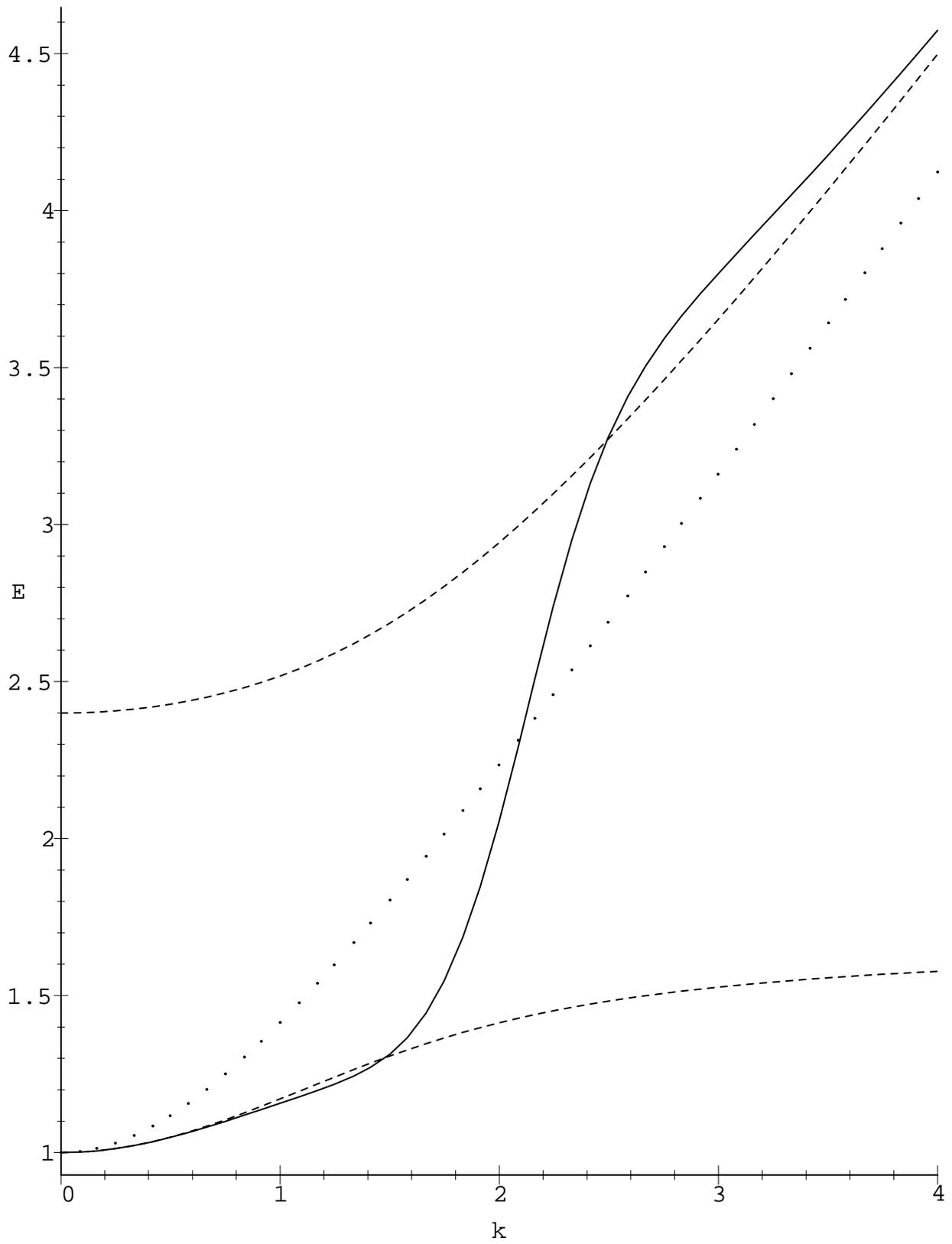
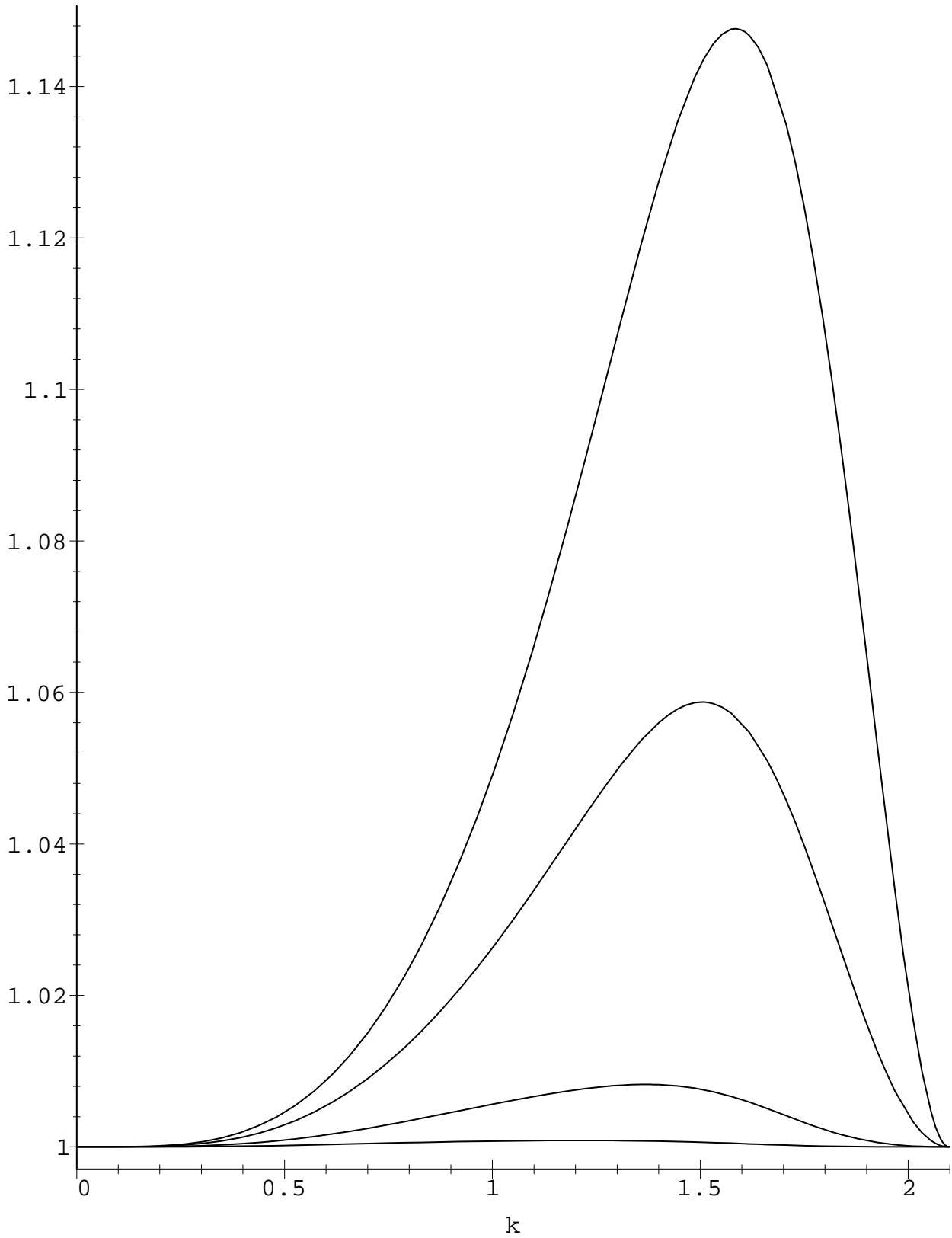
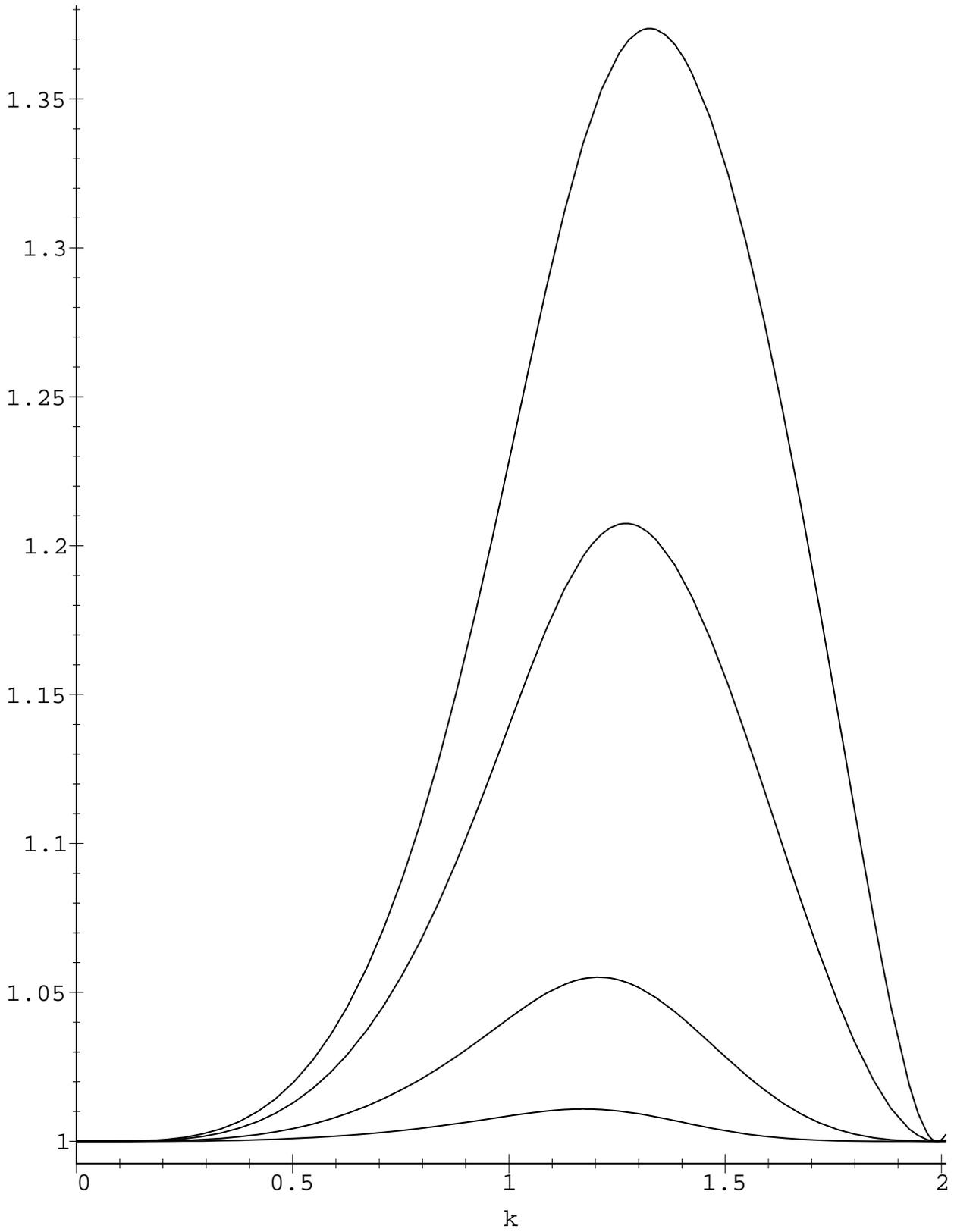
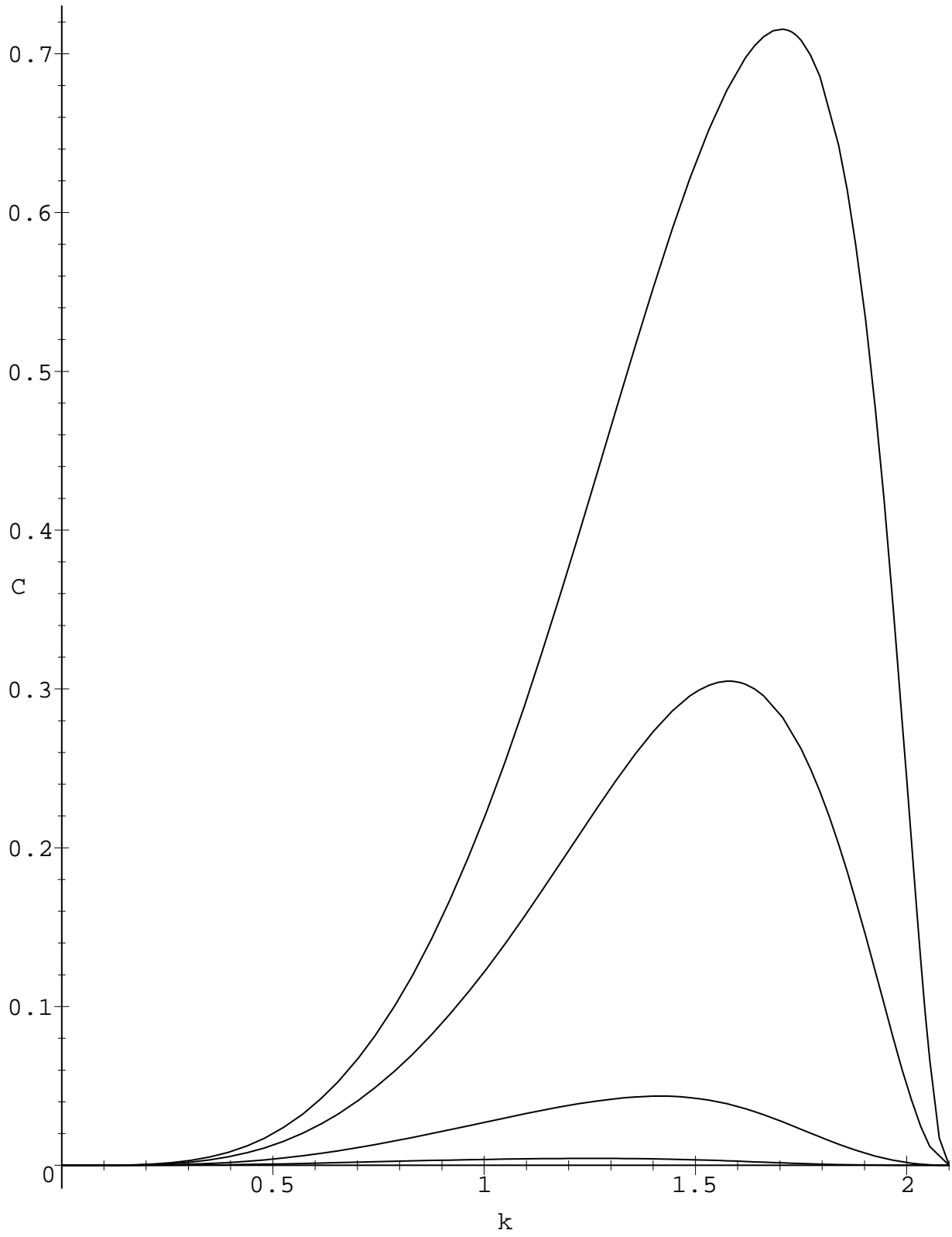
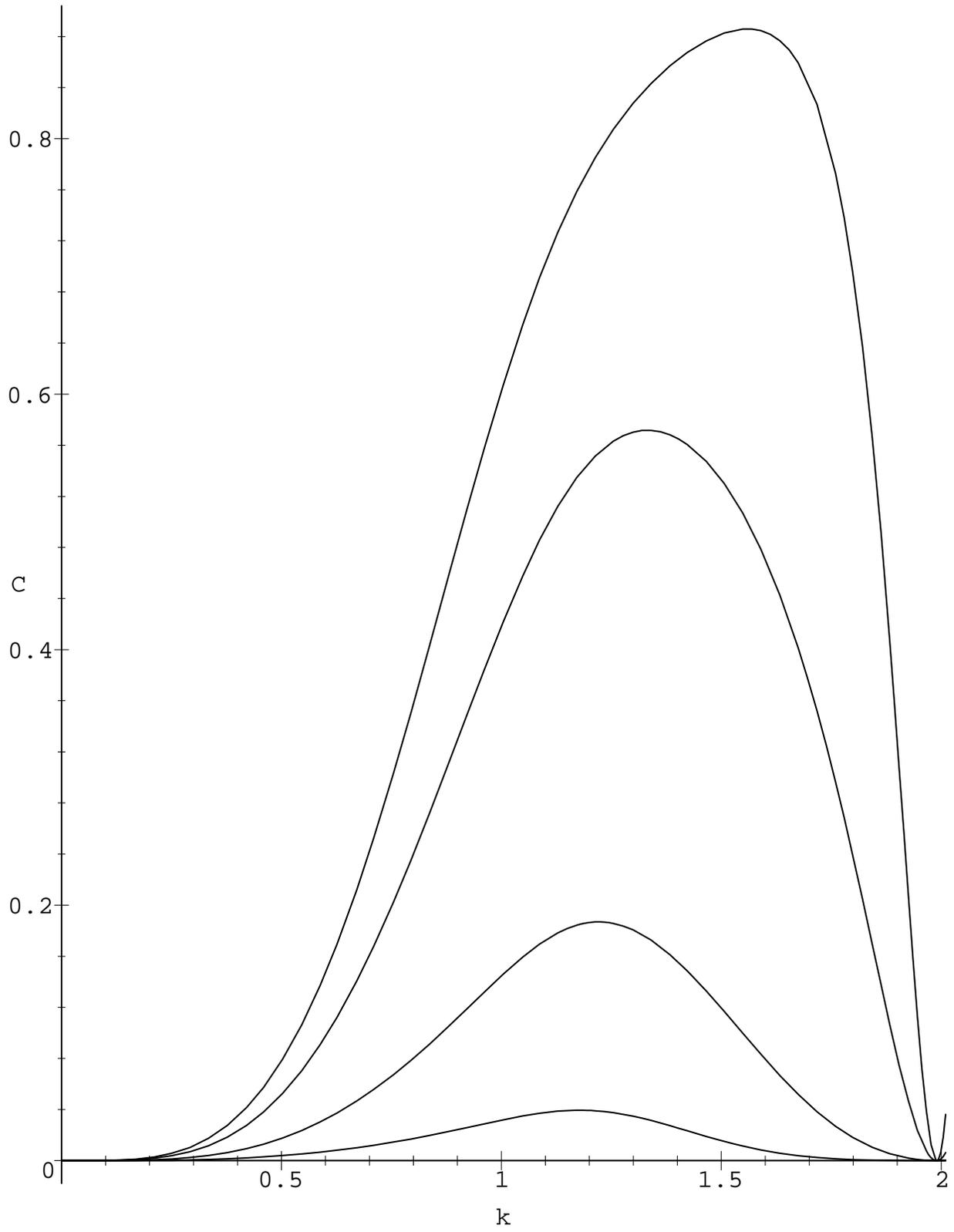
\end{document}